# Interface Ferromagnetism in (110)-Oriented $La_{0.7}Sr_{0.3}MnO_3/SrTiO_3$ Ultrathin Superlattices

J. X. Ma[1], X.F. Liu[1], T. Lin[1], G.Y. Gao[2], J. P. Zhang[3], W.B. Wu[2], X.G. Li[2], and Jing Shi[1]*

1. Department of Physics & Astronomy, University of California,
Riverside, CA92521

2. Hefei National Laboratory for Physical Sciences at Microscale, University of Science and Technology of China, Hefei 230026, China

3. Materials Department, University of California, Santa Barbara, CA 93106-5050

**Abstract**

We explore manganite interface magnetism in epitaxially grown $La_{0.7}Sr_{0.3}MnO_3$(LSMO)/$SrTiO_3$ ultrathin superlattices (SL) along (110) orientation. we show that robust ferromagnetism persists down to four monolayers LSMO(MLs) (~1.1 nm in thickness), of which 50% Mn is at the interface state. Above eight MLs, the magnetic moment is nearly saturated to the theoretical value of 3.7 $\mu_B$, with an estimated interface moment of 3.2 $\mu_B$. In comparison to (100)-oriented SLs which were previously shown to have a spin canted ground state, (110)-oriented SLs exhibit stronger low-dimensional ferromagnetism and better metallicity, suggesting a ferromagnetic interface spin state well suited for all-oxide spintronic devices. The underlining mechanism is qualitatively discussed.

PACS: 75.47.Lx; 75.47.Gk; 75.70.Cn; 75.70.-i

## I. INTRODUCTION

Recent advances in laser molecular beam expitaxy (Laser-MBE) technique with atomic layer controllability have promoted intensive exploration of various superlattices (SL) of perovskite transition metal oxides (TMO). Exotic properties and novel functionalities that are different from their constituent materials or even do not exist in nature have been realized in such artificial SL.[1,2,3,4,5,6] Besides, SL is also a prototype system for studying the hetero-interface and junction properties, interlayers spin coupling, low-dimensional magnetic and transport properties, which are very important for both fundamental physics and technological application.[7,8,9] Among various TMOs, the colossal magnetoresistive and half-metallic manganites[10,11,12] have been extensively studied, due to their intriguing properties and potential of spintronic devices application such as field-effect transistors,[13,14] magnetic tunnel junctions (MTJ),[15,16] and spin valves.[17,18]

So far however, attention has been mainly focused on (100)-oriented SLs, multilayer structures, and hetero-interfaces, probably due to the simplicity of growth.[19,20] It was found that the spin state at (100) interface is rather complex and quite different from the bulk ferromagnetic (FM) spin state. The ferromagnetism of the (100) interface/surface is severely suppressed, which has been attributed to the charge redistribution (modulation) effect as well as the orbital ordering with C-type or A-type antiferromagnetic (AFM) spin interaction induced by strains or broken symmetry.[21,22,23] Modification of the doping profile near the interface can only moderately improve the interface magnetism.[24,25] We note that unlike ordinary itinerant ferromagnets, the spin state of the double-exchange/superexchange mediated manganites may be sensitive to the length and angle of Mn-O-Mn bonds and the local density of orbital states near the interface, both of which are highly orientation-dependent. Furthermore, the $e_g$-orbital occupation can be altered by the anisotropic strains or symmetry-breaking effects which are also orientation dependent. Therefore, orientation may strongly influence the interface spin state.

Considering the crystalline orientation effect on the interface spin state, (110)-oriented interface is particularly interesting. As illustrated in Fig.1, several advantages of (110) interface are anticipated (detailed discussion is presented later): (1) less charge modulation(redistribution); (2) less prone to AFM orbital ordering; (3) more compact layer stacking therefore stronger interlayer spin-spin interaction. Furthermore, as opposed to (100)-oriented $La_{1-x}Sr_xMnO_3$(LSMO)/$SrTiO_3$(STO) SL or multilayer structures, in which the two hetero-interfaces are chemically and therefore electronically different, the two hetero-interfaces in (110)-oriented structures are identical. The (110)-oriented SL with inversion-symmetry may serve as a model system for studying the effects of coupling and charge transfer in TMO, as well as the low-dimensional phenomena in strongly correlated electron systems. Previous studies on (110)-oriented $La_{2/3}Ca_{1/3}MnO_3$ films focused on the magnetic properties and electronic homogeneity of the films.[26,27,28] It is the goal of this work to address the interface spin state of (110)-oriented LSMO/STO.

We synthesized a series of (110)-oriented SL samples consisting of ultrathin FM $La_{0.7}Sr_{0.3}MnO_3$ and non-magnetic insulator STO. By changing the number of atomic layers in LSMO, we effectively tune the interface-to-bulk ratio. We can extract the absolute value of the interface magnetic moment from the SQUID magnetometry measurements on the series of samples. In the following we denote N-monolayers of LSMO in (100) and (110) orientations by "N ML(100)" and "N ML(110)" respectively. We found that robust FM is preserved down to four ML(110) SL, of which 50% Mn ions are at the interface states. Above 8 ML(110), the saturation magnetic moment of Mn is close to the ideal value.

## II. EXPERIMENT

(110)-oriented (LSMO[*N*]/STO[3])$_p$ SL (denoted as N-(110)SL) samples were synthesized with our laser molecular beam epitaxy system.[29] *p* represents the number of LSMO/STO unit cells in the entire SL. STO is fixed at 3 ML(110), whereas *N* is the number of LSMO atomic layers in each SL unit which varies from 3 to 15 ML(110), but the total LSMO in all SL samples, i.e. $N \cdot p$, is kept close to 100 ML(110). As the bulk reference sample, a 100 ML(110) LSMO film capped with a 3 ML(110) STO was grown. Several (100)-oriented SL (denoted as *N*-(100)SL) samples were grown for direct comparison with the (110) SL samples. STO(110) substrates were ultrasound-cleaned in acetone and in-situ annealed in $10^{-6}$ torr ozone atmosphere at 800 ºC for one hour. With this modified substrate treatment,[30] the reflection high energy electron diffraction (RHEED) intensity oscillations were readily obtained. The growth temperature was kept at 750 ºC and the oxygen pressure was 1 mtorr with 12 wt% of ozone. The growth condition was optimized, as indicated by the excellent properties (e.g. the Curie temperature Tc~340 K and the saturation magnetic moment Ms~3.7$\mu_B$) in as-grown 27 nm LSMO film. To ensure sharp interfaces and minimize inter-diffusion in SL growth, interval annealing of 2 to 3 minutes was performed between the growth of LSMO and STO until the RHEED intensity was fully recovered. The procedure was repeated so that the RHEED oscillation intensity was maintained approximately constant throughout the SL growth. Fig.2 shows the typical RHEED intensity oscillations of 7-SL with 14 units. The inset (right) shows the post-growth RHEED pattern, which is almost same as that of the pre-growth, and the Kikuchi lines are clearly visible. The inset (left) shows a typical AFM image of an annealed STO(110) substrate. To further minimize the inter-diffusion upon growth, samples were immediately cooled down to 500 ºC at a quite high rate of 30 ºC/min and then slowly cooled down to room temperature at 10 ºC/min. During cooling, the oxygen pressure was kept same as that during growth. No further post-annealing is conducted. We have found that both *Tc* and magnetic moment of our as-grown samples are nearly unaffected by the post-annealing in oxygen atmosphere of 1 atm pressure at 700 ºC for 10 hours.

The SL structures are characterized by x-ray diffraction (XRD) and high-resolution transmission electron microscopy (HRTEM) recorded with a high-angle annular dark-field detector attached. Magnetic properties of SL samples are measured with SQUID, and transport properties are measured by the standard 4-point method. The magnetic moment is measured during warming in a magnetic field of 0.05 T applied along the in-plane [100] direction (easy axis) upon zero-field cooling. The resistance is measured at zero magnetic field and the resistivity is calculated based on the total number of LSMO layers in SL. The electrical current is along the [100] axis.

## III. RESULTS AND DISCUSSIONS

### A. Structural characterization

Fig.3 shows HRTEM images of 7(110)-SL sample. The bright and dark regions correspond to the LSMO and STO layers, respectively. The epitaxial SL is free of visible impurity phases or structural defects. The interfaces are atomically sharp and smooth, which is crucial for ultrathin SL.[31] Fig.4(a) shows the XRD linear ($\theta$-$2\theta$) scans of 3(110)-SL, 4(110)-SL and 5(110)-SL samples. It is seen that apart from the STO(110) peak from the substrates, satellite reflections from the SL are present, confirming that the SLs have long-range periodicity and good crystallinity. Based on the main and satellite reflections indexed as SL(0), SL(-1), and SL(+1) for each SL, the periods of the 3(110)-SL, 4(110)-SL, and 5(110)-SL are calculated to be 1.64(3), 1.94(1), and 2.16(9) nm, respectively, in agreement with the expected layer thicknesses. Around each SL(0) main Bragg reflection, the Laue fringe is also observable, indicating that the SLs are smooth with uniform thickness, consistent with the HRTEM results. The coherent growth of the SLs on STO(110) substrates is also confirmed by the high-resolution off-specular reciprocal space mapping (RSM) around the (130) reflections. As shown in Fig. 4(b), the sharp main reflection from the 3(110)-SL has the exactly same in-plane $Q_{-110}$ as that of the substrate, indicating coherent epitaxial growth of the SLs on STO(110) substrates. The lattice constants of STO and LSMO are 3.905 Å and 3.876 Å respectively, equivalent to 0.74% tensile strain.

### B. Magnetic and Transport measurements

Figures 5(a) and 5(b) show the temperature dependence of the magnetic moment $M$ and resistivity $\rho$ respectively. Fig. 6 shows magnetic hysteresis loops of SLs measured at 5 K. All SL samples except for 3-SL show well-defined hysteresis loops. For 100-ML(110) LSMO reference film, the saturation moment $M_S$ is 3.75 +/- 0.12 $\mu_B$, which is close to the theoretical value of 3.7 $\mu_B$ expected for bulk material. For samples above 8 ML(110), $M$ reaches saturation at 0.5 T, indicating that $M$ of both inner layers and interfaces is saturated. $M_S$ changes slightly with varying LSMO thickness, and we attribute this to the effect of the interface as the interface/bulk ratio varies. Below 6 ML(110), the moment at 2 T takes a sharp dive as the LSMO thickness decreases. This cannot be accounted for with the interface effect alone. For 3-(110)SL, the remanent moment approaches zero, despite a sizable $M$ of 0.9 $\mu_B$ at 2 T.

In $\rho$-$T$ curves, the resistivity at both room temperature and low temperature increases with decreasing LSMO thickness, which is expected from the enhanced scattering by the interfaces. Metal-insulator transitions are observed for samples above 6 ML(110), but samples below 5 ML(110) show insulating behaviors. It is consistent with the temperature dependent behavior of the magnetic moment. We assign 6 ML(110) the critical thickness, above which SLs are metallic with large *M* and below which are insulating with suppressed *M*.

In manganites, FM spin-spin interaction is mediated by the double-exchange between the nearest $Mn^{3+}$ and $Mn^{4+}$ via oxygen atoms. Away from the LSMO/STO interfaces, charge profile builds up over the scale of the charge screening length in LSMO.[32] Therefore it is reasonable to assume that the magnetic moment arises over the comparable length scale due to the double-exchange mechanism. Kavich et al.[33] studied the interface moment profile for (100)-oriented LSMO films and found that *M* gradually increases and reaches saturation at the 6th layer. Here we adopt a similar profile for (110)-oriented interfaces as illustrated in the inset of Fig.7. Above the critical thickness $N_C$, the inner layers reach saturation at temperatures sufficiently lower than *Tc*. As the LSMO thickness increases further, the interface moment profile remains unchanged; therefore, the total reduction of *M* associated with the interfaces remains constant. Then we obtain: $(M_S^0 - M_S) \cdot N = 2a$, where $M_S^0$ is the saturation moment of inner layer and $M_S$ is the measured average saturation moment at 2 T; *N* is the number of LSMO layers in one SL unit; and *a* is the total moment loss of one interface. Thus, we have $M_S = M_S^0 - 2a/N$. Fig. 7 shows the average saturation moment vs. 1/*N* curve measured at 5 K. It follows a straight line for *N*≥8. By linear fitting, we obtain: $M_S^0$=3.76 μB, and *a*=0.98 μB. $M_S^0$ is very close to the ideal value 3.7 μB expected for thick LSMO films. When LSMO is below 6 ML(110), the data depart sharply from the linear behavior, indicating a crossover thickness of 6 ML(110) below which the entire LSMO plays a role of the interface. If a linear profile is assumed, the magnetic moment starts from 3.2 μB at the outmost interface, which is as high as 87% of inner layer saturation moment, and restores to its full value of 3.7 μB at the fourth layer from the interface.

For direct comparison, we also grew three (100)-oriented (LSMO[N]/STO[2])$_p$ samples with the same thicknesses and periodicities as those of the (110) SL samples. The 70-(100)SL, 7-(100)SL, and 5-(100)SL samples correspond to 100-(110)SL, 10-(110)SL, and 7-(110)SL samples respectively. The magnetic and transport properties are summarized in Table 1. $M\sim T$, $\rho\sim T$, and $M\sim H$ of (100)-oriented SL are also included in Fig. 5 (a) & (b), and Fig.6 with same color as those of the corresponding (110) SL samples. For both 70-(100)SL and 100-(110)SL, the bulk properties are recovered. The FM transition temperatures of (100)-oriented SLs are consistently lower than those of the corresponding (110)-oriented SLs. The moments of 5-(100)SL and 7-(100)SL measured at 2T are about 2.6 $\mu_B$ and 3.1 $\mu_B$, respectively, which are consistent with the reported values in literature.[8, 19, 20] This represents a reduction in *M* per Mn ion (averaged over both inner and interface Mn ions) of 1.15 $\mu_B$ and 0.7 $\mu_B$ for 5-(100)SL and 7-(100)SL samples respectively from the bulk value, which are considerably larger than those of (110) counterparts (i.e. 0.62 $\mu_B$ and 0.13 $\mu_B$ for 7-(110)SL and 10-(110)SL respectively). If a non-linear interface profile is considered, the interface moment would be even smaller for (100) interfaces. Additionally, the remanent moment is about 37% and 28% lower in those (100)-oriented SLs than that in the corresponding (110)-oriented SLs respectively, suggesting different spin states at (100) and (110) interfaces.

Remarkable differences in resistivity are observed between (100)- and (110)-oriented SLs. 5-(100)SL is much more resistive than its (110) counterpart 7-(110)SL, suggesting FM/AFM phase separation[19,20] as a result of interface overdoping. Actually FM metallicity persists down to 6-(110)SL, which is twice as thick as the (110) magnetic interface obtained from the preceding analysis. Interestingly, the resistivity ratio $\rho(100)/\rho(110)$ is nearly constant below $T_c$, as shown in the inset of Fig. 2(b). This ratio is 100 and 5 for 5-(100)SL/7-(110)SL and 7-(100)SL/10-

(110)SL respectively. These facts confirm that (110)-oriented interface has more robust ferromagnetism and metallicity than (100)-oriented interface.

Here we qualitatively discuss the possible underlying mechanism of the observed strong ferromagnetism of (110) interface. It is noted that the (100) interface is 6 ML(100) thick (~ 2.4 nm), nearly an order of magnitude larger than the Thomas-Fermi screening length $L_{TF}$ of 0.3 nm of LSMO at this doping level, and the interface magnetic moment is only 40% of the bulk value.[33] In stark contrast to (100) interface, the magnetic (110) interface is 3 ML(110) thick (~0.8 nm), only about three times as along as $L_{TF}$~0.3 nm, but the magnetic moment is as high as 87% of the bulk value, which is even greater than that of LaMnO$_3$ modified interface (~80%).[33] It suggests that the charge density modulation alone may account for the moment reduction of (110) interface. On the contrary, for (100) interface, other mechanisms than the charge density modulation must play a decisive role in the suppressed interface magnetism.

Now let us discuss the state of Mn ions near the interface. As shown in Fig. 1, the red squares indicate the unit cell at the interfaces. For (100)-oriented structures, the two interfaces are different. Each Mn ion is surrounded by a La/Sr ion for the upper interface and surrounded by ½ La/Sr and ½ Sr for the lower interface. Kumigashira *et al.* reported that all Ti ions in STO are in Ti$^{4+}$ state, ruling out charge transfer from Mn to Ti.[34] Thus, the lower interface is over-doped. For (110) interface however, the Mn ion is surrounded by ¾ La/Sr ions and ¼ Sr ions. Thus, the Mn ion at (110) interface is 50% less over-doped compared to that at (100) MnO$_2$-SrO-TiO$_2$ interface. Qualitatively, the effective doping at (110) interface is still under 0.5, whereas it is over 0.5 at (100) MnO$_2$-SrO-TiO$_2$ interface, which places the latter in the AFM regime.[21,35] We argued earlier this over-doping related charge profile alone cannot explain the large moment reduction in (100) interface. We believe that the second mechanism is related to the orbital ordering effect. It is known that the orbital ordering with C- or A-type AFM is caused by the preferential occupation of $d_{z^2}$ or $d_{x^2-y^2}$ orbitals in the presence of symmetry breaking or strain at the (100)-oriented interface/surface of LSMO/STO.[21,22,23] For (110) interface, the crystal field variation due to the strain effects or symmetry-breaking is along [110] direction, which does not split $d_{z^2}$ and $d_{x^2-y^2}$ orbital states as it does in [100] direction. This may lessen the orbital ordering and consequently AFM interaction at (110)-oriented interface/surface. Thirdly, there are two oxygen atoms between the adjacent (110) layers but only one between the adjacent (100) layers; therefore, the interlayer Mn-O-Mn double-exchange coupling is stronger for (110), which results in a more rapid recovery to the bulk spin state. Note that the FM properties of (110) interface may be further improved by using similar strategy proposed by Yamada et al. based on additional charge profile modification.[23]

## IV. CONCLUSION

High quality epitaxial LSMO/STO SLs with ultrathin LSMO have been successfully fabricated in (110) orientation. In comparison to (100)-oriented SLs grown in the same conditions, (110)-oriented SLs show more favorable low-dimensional ferromagnetism and metallicity, which are important for spintronic applications. The underlining mechanism has been qualitatively discussed. Further experiments on the orientation-dependence of orbital states by linear dichroism in x-ray absorption spectroscopy are under way.

**ACKNOWLEDGEMENT**

Work at UCR is supported by DMEA/CNN under award H94003-08-2-0803 and NSF ECCS-0802214.

**Figure Captions**

Fig.1. (Color online) Schematic view of LSMO/STO/LSMO hetero-interfaces. The upper interface LaSr-O/MnO$_2$/Sr-O is different from the lower interface LaSr-O/MnO$_2$/LaSr-O for (100) orientation. Both interfaces identical: O$_2$/LaSrMnO/O$_2$ for (110) orientation.

Fig.2 Typical RHEED intensity oscillations for LSMO/STO superlattices growth. Insets show the post-growth RHEED pattern (right) and AFM image of annealed STO substrate (left).

Fig. 3 Z-contrast TEM image of 7-SL samples with 14 periodic units in full range. Inset is the high-resolution image taken near the substrate. The bright and dark regions correspond to LSMO and STO, respectively.

Fig.4. (a) XRD linear scans from three SL samples, 3-SL, 4-SL, and 5-SL as denoted. Aside from the STO(110) reflections, those from the SLs are indexed as SL(-1), SL(0), and SL(+1), respectively. (b) XRD RSM on (130) reflections from the 3-SL sample. Note that the SL(130) main reflection has the same in-plane Q value as that of STO(130), indicating a coherent growth of the SLs.

Fig. 5. (Color online) (a) Temperature dependence of magnetic moment *M* for SL samples measured during field warming (0.05T along [001]) upon zero field cooling (b) Temperature dependence of resistivity ρ measured at zero magnetic field. Resistivity is normalized by the total number of LSMO layers. Inset shows the resistivity ratio of (100)- and (110)-oriented SL samples vs. *T* curves.

Fig. 6. (Color online) Magnetic hysteresis loops of SL samples measured at 5 K.

Fig. 7. (Color online) Saturation magnetic moments at 2 T vs. *1/N*, where *N* is the number of LSMO SL layers in a SL unit. The red line represents linear fitting $M_S = M_S^0 - 2a/N$. Inset shows schematic profiles of the magnetic moment of SL samples. Above the critical thickness of 8 MLs, the inner layer moment is saturated to $M_S^0$ and the total loss of the interface moment remains fixed as the LSMO thickness varies. Below 8 MLs, the magnetic moments of both the inner layers and the interface layers decrease with decreasing thickness.

Table 1. Direct comparison of magnetic and transport properties between (100)- and (110)-oriented SL samples with same thickness and periodicities. The overall properties of (110)-oriented SLs are superior to that of (100)-oriented SLs.

|  | $M_s$ ($\mu B$) | $M_0$ ($\mu B$) | $T_c$ (K) | $T_p$ (K) | $\rho$ ($\Omega cm$) |
|---|---|---|---|---|---|
| 7-(110)SL | 3.08 | 2.2 | 180 | 164 | 0.059 |
| 5-(100)SL | 2.55 | 1.38 | 165 | <85 | 0.23 |
| 10-(110)SL | 3.57 | 2.5 | 265 | 285 | 0.024 |
| 7-(100)SL | 3.0 | 1.8 | 245 | 258 | 0.050 |
| 100-(110)SL | 3.75 | 3.3 | 340 | >Tc | 0.0033 |
| 70-(100)SL | 3.76 | 2.1 | 330 | >Tc | 0.0044 |

Ma *et al.* Table 1